\documentstyle[twoside,fleqn,espcrc2,epsf]{article}

\title{
  \vspace{-2cm}
  \begin{flushright}
    KANAZAWA 99-20\ \ \ \ \ \ \\
    August 1999\ \ \ \ \ \ \  
  \end{flushright}
  \vspace*{-.5cm}
  A perfect monopole action for SU(2) QCD
  \thanks{
    Presented by S.Fujimoto
  }
}

\author{
  Shouji Fujimoto,
  Seikou Kato
  and Tsuneo Suzuki \\
\vspace{3mm}
  Institute for Theoretical Physics, Kanazawa University,
  Kanazawa 920-1192, Japan
}

\begin{document}

\begin{abstract}
We found a quantum perfect lattice action in the 4-dimensional
monopole current theory which is known as an effective theory in the
infrared region of QCD.
The perfect monopole action is transformed exactly into a lattice
action of a string model.
When the monopole interactions are weak as in the case of infrared
SU(2) QCD, the string interactions are strong.
The static potential and the string tension in this region can be
estimated analytically by the use of the strong coupling expansion.
\end{abstract}

\maketitle

\section{Introduction}
We consider the following steps to obtain the renormalized trajectory as
the continuum limit of lattice field theory.
(I)Performing $n$ steps of block spin transformations from a-lattice
to b-lattice.
Here a-lattice is the lattice whose lattice constant is $a$, and
b-lattice has the lattice constant $b=na$.
(II)In order to go to the continuum limit keeping physical scale $b$
finite, we take the limit $a\to 0$ and $n\to\infty$ for fixed $b$.
(III)One can find the perfect action and the perfect operator which
reproduce the  continuum value on the b-lattice.

In general, however, the above scenario is difficult to carry out.
We suggest a simple but a non-trivial lattice model composed of
monopole two-point interactions alone to derive analytically the
renormalized trajectory and the perfect operator corresponding to a
potential between static electric charges.
This is similar to the blocking from the continuum theory as developed 
by Bietenholz and Wiese\cite{wiese96}.

The continuum rotational invariance is shown exactly with the operator.
In addition, this model is very interesting, since the effective
monopole action obtained after an abelian projection of pure $SU(2)$
lattice QCD is known to be well dominated by such two-point monopole
interactions alone in the infrared region
\cite{shiba2,shiba3,kato97,kato98}.

\section{
  Perfect lattice action and operator for the monopole current theory
}

The renormalization group flow can be studied for arbitrary
two-point interactions,
but for simplicity we assume the following form of lattice monopole
action.
\[
S[k]=\sum_{s,s',\mu}k_\mu(s)D_0(s-s')k_\mu(s'),
\]
\begin{equation}
D_0(s-s')\!=\!
\beta\Delta^{-1}(s-s')+\alpha\delta_{s,s'}+
\gamma\Delta(s-s').
\label{eqn.D}
\end{equation}

The Wilson loop operator in the monopole current theory is given by
the following operator\cite{kato99,stack,misha1,shiba1} on the
a-lattice:
\[
W_m({\cal C})=
\exp\left\{
  2\pi i\sum_{s,\mu}N_{\mu}(s,S^J)k_{\mu}(s)
\right\},
\]
\[
N_{\mu}(s,S_J\!)\equiv\!\!\sum_{s'}\Delta^{{}^{-1}}\!\!(s-s')\frac{1}{2}
\epsilon_{\mu\alpha\beta\gamma}\partial_{\alpha}
S^J_{\beta\gamma}(s'+\hat{\mu}),
\]
\begin{equation}
\partial'_{\beta}S^J_{\beta\gamma}(s)\equiv J_{\gamma}(s).
\label{eqn.S}
\end{equation}

Let us construct a blocked monopole current $K_\mu(s)$ on the coarse
b-lattice satisfying the conservation law:
\begin{eqnarray}
K_\mu(s)\!\!\!\!&=&\!\!\!\!\!\!
  \sum_{i,j,l=0}^{n-1}\!\!
    k_\mu(ns+(n-1)\hat{\mu}+i\hat{\nu}+j\hat{\rho}+l\hat{\sigma})
\nonumber \\
&\equiv&{\cal B}_{k_\mu}(s).
\nonumber
\end{eqnarray}
It satisfies the conservation law $\sum_\mu\partial'_\mu K_\mu(s)=0$.

The vacuum expectation value of the Wilson loop is
\[
\langle W_m({\cal C}) \rangle =
  \!\!
  \sum_{K_{\mu}=-\infty\atop{\partial^{\prime}_{\mu}K_{\mu}=0}}^{\infty}
  \!\!
  Z[K,J]
  \Bigg/
  \!\!
  \sum_{K_{\mu}=-\infty\atop{\partial^{\prime}_{\mu}K_{\mu}=0}}^{\infty}
  \!\!
  Z[K,0],
\]
where
\begin{eqnarray}
&&\!\!\!\!\!\!\!\!\!\!
Z[K,J] =
  \!\!\!\!
  \sum_{k_{\mu}=-\infty\atop{\partial^{\prime}_{\mu}k_{\mu}=0}}^{\infty}
  \!\!\!\!
  \exp
  \Bigg\{
    -\!\!\sum_{s,s',\mu}\!\! k_{\mu}(s)D_0(s-s')k_{\mu}(s')
\nonumber \\
&&\!\!\!\!\!\!\!\!\!\!
    +2\pi i\sum_{s,\mu} N_{\mu}(s)k_{\mu}(s)
  \Bigg\}
  \delta
      \left(
        K_{\mu}(s^{(n)})
        -{\cal B}_{k_\mu}(s^{(n)})
      \right).
\nonumber
\end{eqnarray}
Introducing auxiliary fields $\phi$ and $\gamma$, we rewrite the constraints
$\partial^{\prime}_{\mu}k_{\mu}=0$ and
$K_{\mu}(s^{(n)})={\cal B}_{k_\mu}(s^{(n)})$.
Then we change the integral region of $\gamma$ and $\phi$
from the first Brilluin zone to the infinite region, since the monopole
currents take integer values.
Making use of the Poisson sum rule and recovering dimensional lattice
constants $a$ and $b=na$,
we carry out explicitly integrals with respect to $F$, $\phi$, $\gamma$
and take the continuum limit($a\to 0,n\to\infty,$ for fixed $b=na$).
We obtain the perfect action and the perfect operator
\begin{eqnarray}
&&\!\!\!\!\!\!\!\!\!\!
\langle W_m({\cal C}) \rangle
\!\!
=
\!\!
  \exp\left\{
      {\cal S}_{\bf 1}[N]
    + {\cal S}_{\bf 2}[B]
  \right\}
\nonumber \\
&&
  \times
  \frac{\displaystyle
    \sum_{b^3K_\mu(bs)=-\infty\atop\partial'_\mu K_\mu=0}^{\infty}
    \!\!\!\!\!\!\!\!
    \exp\left\{
        {\cal S}_{\bf 3}[K]
      + {\cal S}_{\bf 4}[K,B]
    \right\}
  }{\displaystyle
    \sum_{b^3K_\mu(bs)=-\infty\atop\partial'_\mu K_\mu=0}^{\infty}
    \!\!\!\!\!\!
    \exp\left\{
        {\cal S}_{\bf 3}[K]
    \right\}
  },
\nonumber
\end{eqnarray}
where ${\cal S}_{\bf 1}[N]$ is an interaction term between surface
source $N_\mu(x)$ in the continuum space time
\begin{equation}
  - \pi^2 \!\!\int_{-\infty}^{\infty}\!\!\!\!d^4xd^4y
  \sum_{\mu}N_{\mu}(x)D_0^{-1}(x-y)N_{\mu}(y).
\label{S1:mon}
\end{equation}
${\cal S}_{\bf 2}[B]$ is an interaction term between the surface source
$B_\mu(s)$ on the coarse lattice:
\[
  \pi^2 b^8 \sum_{s,s'}\sum_{\mu,\nu}
  B_{\mu}(bs)
    D_{\mu\nu}(bs-bs')
  B_{\nu}(bs'),
\]
\begin{eqnarray}
&&\!\!\!\!\!\!\!\!\!\!\!\!
B_\mu(bs^{(n)})\equiv
  \lim_{a\to 0 \atop{n\to\infty}}
  a^8\sum_{s,s',\nu}
    \Pi_{\neg\nu}(bs^{(n)}-as)
\nonumber \\
&&
\times
  \left\{
    \delta_{\mu\nu}
    -\frac{\partial_{\mu}\partial'_{\nu}}{\sum_{\rho}\partial_{\rho}
     \partial'_{\rho}}
  \right\}
  D_0^{-1}(as-as')N_{\nu}(as').
\nonumber
\end{eqnarray}
Here $\Pi_{\neg\mu}(nas^{(n)}-as)$ is defined as follows:
\begin{eqnarray}
&&\!\!\!\!\!\!\!\!\!\!
\Pi_{\neg\mu}(nas^{(n)}-as)
\equiv
\nonumber \\
&&
  \frac{1}{n^3}
  \delta\left( nas_\mu^{(n)}+(n-1)a-as_\mu \right)
\nonumber \\
&&
\times
  \prod_{i(\ne \mu)}\left(
    \sum_{I=0}^{n-1}\delta\left( nas_i^{(n)}+Ia-as_i \right)
  \right).
\nonumber
\end{eqnarray}
${\cal S}_{\bf 3}[K]$ is a perfect monopole action on the coarse lattice
\[
  - b^8 \sum_{s,s'}\sum_{\mu,\nu}
  K_{\mu}(bs)
    D_{\mu\nu}(bs-bs')
  K_{\nu}(bs').
\]
${\cal S}_{\bf 4}[K,B]$ is an interaction term between the monopole
current and the modified surface source on the b-lattice.
\[
  2 \pi i b^8 \sum_{s,s'}\sum_{\mu,\nu}
  B_{\mu}(bs)
    D_{\mu\nu}(bs-bs')
  K_{\nu}(bs').
\]
$D_{\mu\nu}(bs^{(n)}-bs^{(n)'})$ is the after some gauge fixing
inverse of the following operator:
\begin{eqnarray}
&&\!\!\!\!\!\!\!\!\!\!
A'_{\mu\nu}(bs^{(n)}-bs^{(n)'})
\equiv
\nonumber \\
&&\!\!\!\!\!\!\!\!
\lim_{a\to 0\atop{n\to \infty}}
  a^8\sum_{s,s'}
  \Pi_{\neg\mu}(nas^{(n)}-as)
  \Pi_{\neg\nu}(nas^{(n)'}-as')
\nonumber \\
&&\qquad
  \times
  \left\{
    \delta_{\mu\nu}
    -\frac{\partial_{\mu}\partial'_{\nu}}{\sum_{\rho}\partial_{\rho}
     \partial'_{\rho}}
  \right\}
  D_0^{-1}(as-as').
\nonumber
\end{eqnarray}

\section{String representation of the monopole action}
The quadratic monopole action can be transformed exactly into that of
the string model using the BKT transformation and the dual
transformation \cite{Bere,KT,misha93,maxim98}:
\begin{eqnarray}
&&\!\!\!\!\!\!\!\!\!\!\!\!
\langle W_m({\cal C}) \rangle
=
\exp\left\{ {\cal S}_{\bf 1}[N] \right\}
\!\!\!\!\!\!
\nonumber \\
&&\!\!\!\!\!\!\!\!\!\!
\times
\!\!\!\!\!\!\!\!
\sum_{\sigma_{\mu\nu}(s)=-\infty
  \atop{\partial_{[\alpha}\sigma_{\mu\nu]}(s)=0}}^{\infty}
\!\!\!\!\!\!\!\!\!\!
\exp\left\{
    {\cal S}_{\bf 5}[\sigma]
  + {\cal S}_{\bf 6}[\sigma,B]
\right\}
\!\!\!\!
\Bigg/
\!\!\!\!\!\!\!\!
\sum_{\sigma_{\mu\nu}(s)=-\infty
  \atop{\partial_{[\alpha}\sigma_{\mu\nu]}(s)=0}}^{\infty}
\!\!\!\!\!\!\!\!
\exp\left\{
    {\cal S}_{\bf 5}[\sigma]
\right\}.
\nonumber
\end{eqnarray}
Here ${\cal S}_{\bf 1}[N]$ is the same as in Eq.(\ref{S1:mon}).
${\cal S}_{\bf 5}[\sigma]$ is a perfect action written by string
variables $\sigma_{\mu\nu}(s)$:
\[
- \pi^2\sum_{s,s'}\sum_{\atop{\mu\neq\alpha \atop{\nu\neq\beta}}}
\sigma_{\mu\alpha}(s)H_{\mu\alpha;\nu\beta}(s-s')\sigma_{\nu\beta}(s'),
\]
where $H_{\mu\alpha;\nu\beta}(s-s')$ is
\[
  \partial_{\alpha}\partial_{\beta}'
  \sum_{s_1}D_{\mu\nu}^{-1}(s-s_1)\Delta_{\rm L}^{-2}(s_1-s').
\]
${\cal S}_{\bf 6}[\sigma,B]$ is the interaction term between
$\sigma_{\mu\nu}(s)$ and $B_\mu(s)$:
\[
  - 2\pi^2\sum_{s,s'}\sum_{\mu,\nu}
    \sigma_{\mu\nu}(s)\partial_{\mu}'
    \Delta_{\bf L}^{-1}(s-s')B_{\nu}(s').
\]

Let us assume that the monopole action on the dual lattice 
is in the weak coupling region for large $b$ as 
realized in the infrared region of pure $SU(2)$ QCD. Then the string
model on the original lattice 
is in the strong coupling region. The strong coupling expansion on the
lattice can be performed easily and quantum fluctuations terms which
include more plaquettes become small\cite{kato98,kato99}.

Then the vacuum expectation value for the Wilson loop is evaluated
easily by using only the classical part of the above equation,
that is, $\exp\left\{ {\cal S}_{\bf 1}[N] \right\}$.

\section{The rotational invariance}
The static potential and the string tension can be calculated
analytically.
The plaquette variable $S_{\alpha\beta}$ in Eq.(\ref{eqn.S}) for the 
static potential $V(bI,0,0)$ is expressed by
\begin{eqnarray}
&&\!\!\!\!\!\!\!\!\!\!\!\!
S_{\alpha\beta}(z)=
\nonumber \\
&&\!\!\!\!\!\!\!\!\!\!\!\!
  \delta_{\alpha 1}\delta_{\beta 4}\delta(z_{2})\delta(z_{3})
  \theta(z_{1})\theta(Ib\! -\! z_{1})
  \theta(z_{4})\theta(Tb\! -\! z_{4}).
\nonumber
\end{eqnarray}
Also the variable $S_{\alpha\beta}$ for the static potential
$V(bI,Ib,0)$ is given by
\begin{eqnarray}
&&\!\!\!\!\!\!\!\!\!\!\!\!
S_{\alpha\beta}(z)
=
  \Bigl(
    \delta_{\alpha 1}\delta_{\beta 4}+\delta_{\alpha 2}\delta_{\beta 4}
  \Bigr)
  \delta(z_{3})\theta(z_{4})\theta(Tb-z_{4})
\nonumber \\
&&\!\!\!\!\!\!\!\!\!\!
\times
  \theta(z_{1})\theta(Ib-z_{1})
  \theta(z_{2})\theta(Ib-z_{2})
  \delta(z_{1}-z_{2}).
\nonumber
\end{eqnarray}
Then the static potentials $V(Ib,0,0)$ and $V(Ib,Ib,0)$ can be written
as
\begin{eqnarray}
V(Ib,0,0) &=&
\frac{\pi\kappa Ib}{2} \ln\frac{m_1}{m_2},
\nonumber \\
V(Ib,Ib,0) &=& \frac{\sqrt{2}\pi\kappa Ib}{2} \ln\frac{m_1}{m_2}, 
\nonumber
\end{eqnarray}
where $m_1$ and $m_2$ are expressed by the original couplings in
Eq.(\ref{eqn.D}) as
$\kappa(m_1^2-m_2^2)=1/\gamma$, $m_1^2+m_2^2=\alpha/\gamma$ and
$m_1^2m_2^2=\beta/\gamma$.
The potential takes only the linear form and the rotational invariance
is recovered completely even for the nearest $I=1$ sites.
The string tension is evaluated as
$\sigma=\pi\kappa\ln(m_1/m_2)/2$. 
This is consistent with the analytical results\cite{suzu89} in the
Type-2 superconductor.
The two constants $m_1$ and $m_2$ may be regarded as the coherence and
the penetration lengths.

T.S. acknowledges financial support from JSPS Grant-in Aid for Scientific  
Research (B) (No.10440073 and No.11695029).


\end{document}